\documentclass[sigconf, nonacm]{acmart}

\setcopyright{rightsretained}
\begin{document}
\fancyhead{}

\title{Model-Parallel Model Selection for Deep Learning Systems}

\author{Kabir Nagrecha}
\email{knagrech@ucsd.edu}
\affiliation{%
  \institution{University of California, San Diego}
  \streetaddress{9500 Gilman Dr.}
  \city{La Jolla}
  \state{California}
  \country{USA}
  \postcode{92093}
}

\maketitle
\section{ABSTRACT}

As deep learning becomes more expensive, both in terms of time and compute, inefficiencies in machine learning training prevent practical usage of state-of-the-art models for most users. The newest model architectures are simply too large to be fit onto a single processor. To address the issue, many ML practitioners have turned to model parallelism as a method of distributing the computational requirements across several devices. Unfortunately, the sequential nature of neural networks causes very low efficiency and device utilization in model parallel training jobs.

 \par We propose a new form of “shard parallelism” combining task parallelism and model parallelism, and package it into a framework we name \textsc{Hydra}. \textsc{Hydra} recasts the problem of model parallelism in the multi-model context to produce a fine-grained parallel workload of independent model shards, rather than independent models. This new parallel design promises dramatic speedups relative to the traditional model parallelism paradigm. \par

\section{INTRODUCTION}
The computational costs of deep learning (DL) have grown exponentially over the years, and recent advances in neural network architectures have only continued this trend. Systems such as BERT \cite{devlin2019bert} have pushed the boundaries of accuracy in applied domains. But the memory and computational power required for training such models can be staggering. As a result, single-device model training solutions have become increasingly impractical. Model parallelism has come to the fore as a potential solution. \par
The intuition behind model parallelism is simple: divide the model into shards, which can then be handled on separate devices --- thus reducing the per-device memory footprint to manageable levels. 
\par \textbf{Motivation.}
This form of model parallelism permits massive models to be trained across multiple devices, but it is highly inefficient. As Figure 1 demonstrates, neural network forward inference and backpropagation are inherently sequential tasks with features and gradients being sent from layer to layer. Sharding the model does not change this reality, and as a result, devices in a model-parallel setup must wait idle in every pass until their shard receives the necessary features or gradients. Allowing for the training of massive models is insufficient --- to democratize these state-of-the-art architectures, training must also be efficient and feasible for practitioners.\par

\begin{figure}[h]
  \centering
  \includegraphics[width=2.6in ]{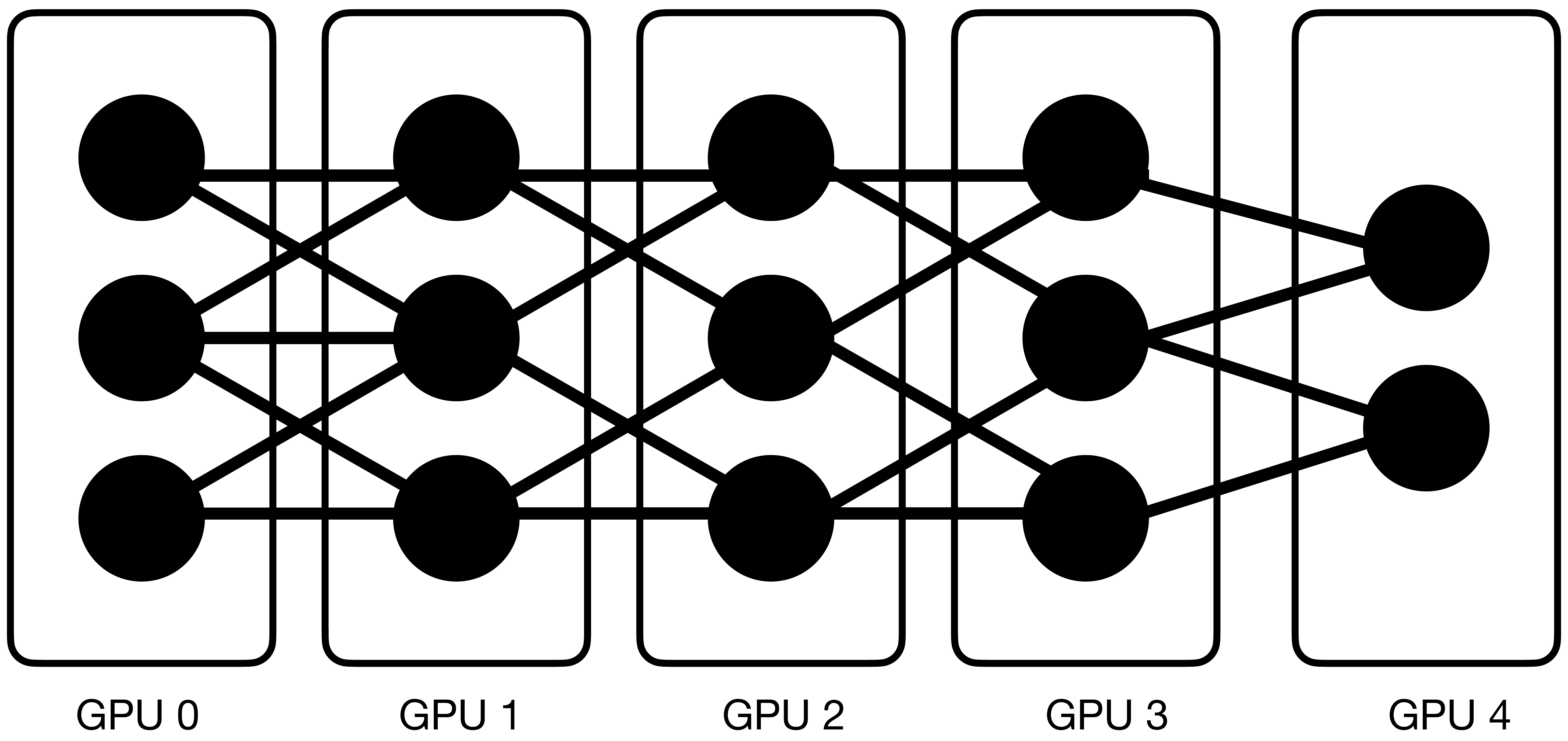}
  \caption{Sharding of a model across 5 devices. Each device must be able to communicate with its immediate neighbors to pass information between shards both for forward inference and backpropagation.}
  \Description{Model shards are placed on different devices with arrows depicting communication between them.}
\end{figure}

Data parallel systems can rely on asynchronous updates to maximize device utilization, as stochastic gradient descent is generally robust to inconsistent data ordering. But there is no analogue for traditional model parallelism to work in a similarly efficient manner.

\par \textbf{Model Selection Integration.}
Let us instead consider the problem in the context of model selection, the process wherein a DL practitioner tests different model configurations to select the best one. Consider, for example, a radiologist building a computer vision system to analyze X-ray scans. They may wish to compare dozens of models or hyper-parameter configurations for the problem. Such training tasks offer an embarrassingly parallel workload. \par We observe that it is possible to leverage this second level of parallelism to solve model parallelism's problem of device underutilization. While one model shard may be untrainable due to its dependencies, an idle device can work operation on a viable shard belonging to an entirely different model. We combine model selection's task parallelism with model parallelism to not only train independent models in parallel, but to train independent model shards in parallel for a fine-grained parallel workload. This ''shard parallelism'' forms the core of \textsc{Hydra}: our proposed framework for model-parallel training in the multiple-model context.

\section{APPROACH}
The core idea behind \textsc{Hydra}'s model-parallel model selection is recasting the training workload from running models in parallel to running shards in parallel. By combining task parallelism with model parallelism to produce a more granular workload, we extend the capacity of model-selection systems to larger-than-memory models, while also solving the problem of model-parallel device underutilization.

\par \textbf{Desiderata.}
To judge our system, we define a list of desiderata. \par
\begin{itemize}
\item A desirable model-parallel model selection system must maximize device utilization. Our approach should theoretically keep all devices utilized throughout the process.
\item Increased model training throughput. Shard parallelism should be capable of improving training throughput relative to either model or task parallelism alone as it offers a more granular parallel workload.
\item Exact replication of model training output. Sharding the model should not affect the integrity of its gradient updates or data transforms in any way.
\end{itemize}
 
 \begin{figure}[hbt!]
  \centering
  \includegraphics[width=3in]{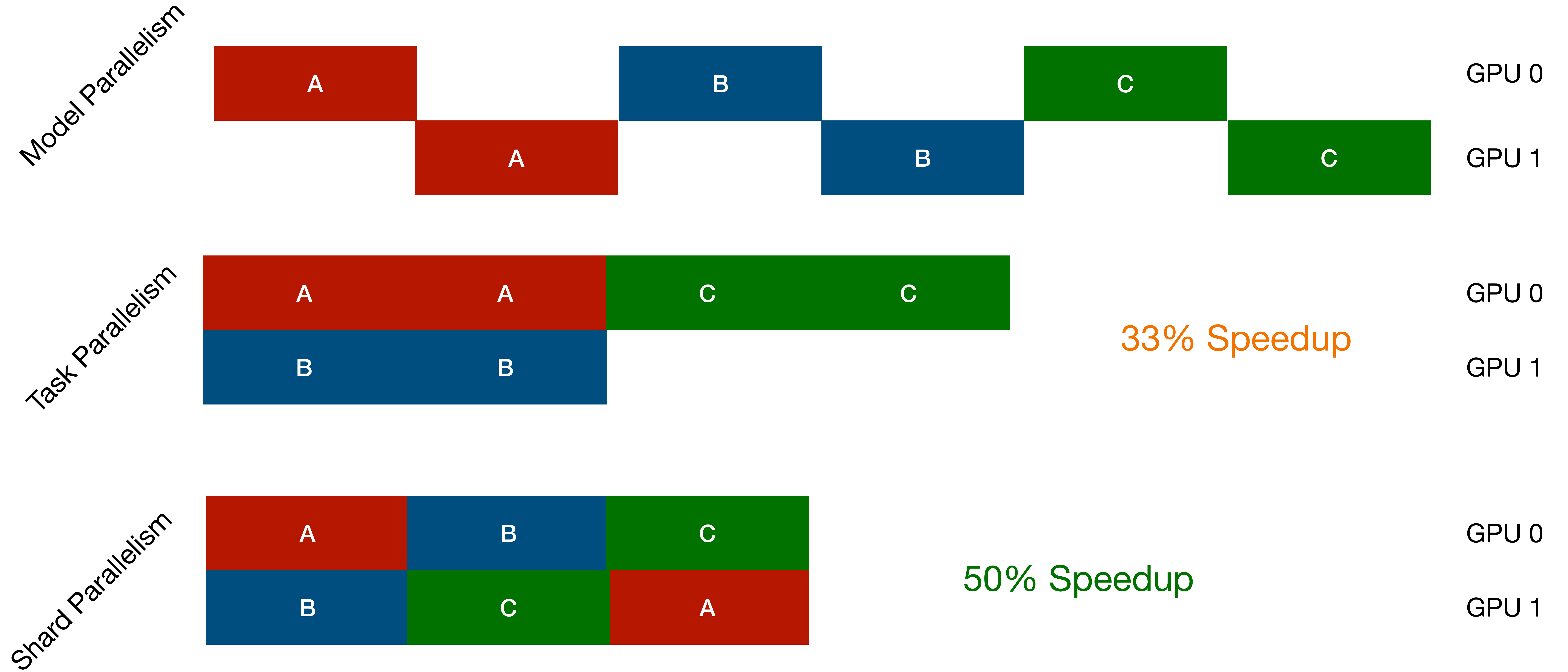}
  \caption{Even with the assumption that the models fit in memory, shard parallelism is substantially more efficient than model parallelism or task parallelism alone in training multiple models.}
  \Description{Shard paralellism is compared to task parallelism and model parallelism in a runtime chart.}
\end{figure}

\par \textbf{Technical Novelty.} Sharded training is an area of research that has received little attention, as most model training systems work in either a task-parallel or model-parallel regime. Operating at a sub-model level is a heretofore unexplored approach, combining the two paradigms to boost efficiency and trainability in a new way.

\section{EVALUATION}
\textbf{Workloads.} We evaluate \textsc{Hydra} on two model architectures to test against our criteria. We use a 1.2 million parameter feedforward neural network to check that \textsc{Hydra} does not harm model accuracy. Because this model is small enough to fit onto a single device's memory, it works well for comparisons between model-parallel results and single-device results. To simulate real-world workloads, we run BERT-Large fine-tuning \cite{turc2019} for 3 epochs on the SQuAD \cite{rajpurkar2016squad} workload. We test on a single cluster with 4 16GB Tesla V100s. Such a configuration is reasonably representative of real-world hardware setups. \par
\textbf{Goals.} Our primary aims are to demonstrate that \textsc{Hydra} meets the aforementioned desiderata, and show its efficiency in comparison to existing model-parallel or task-parallel frameworks.

\subsection{Execution Plan.}
Figure 3 describes \textsc{Hydra}'s architecture. Thus far we have been able to integrate basic multiple-model training with model sharding, forming the core of our shard-parallelism implementation. The next step is the construction of the scheduler to automate the sharding and training of user-specified models. For the final model selection integration, we use Cerebro \cite{inproceedings} as our target system. Cerebro's use of data parallelism offers an additional level of optimization that \textsc{Hydra} can leverage for efficient training.

 \begin{figure}
  \centering
  \includegraphics[width=3in]{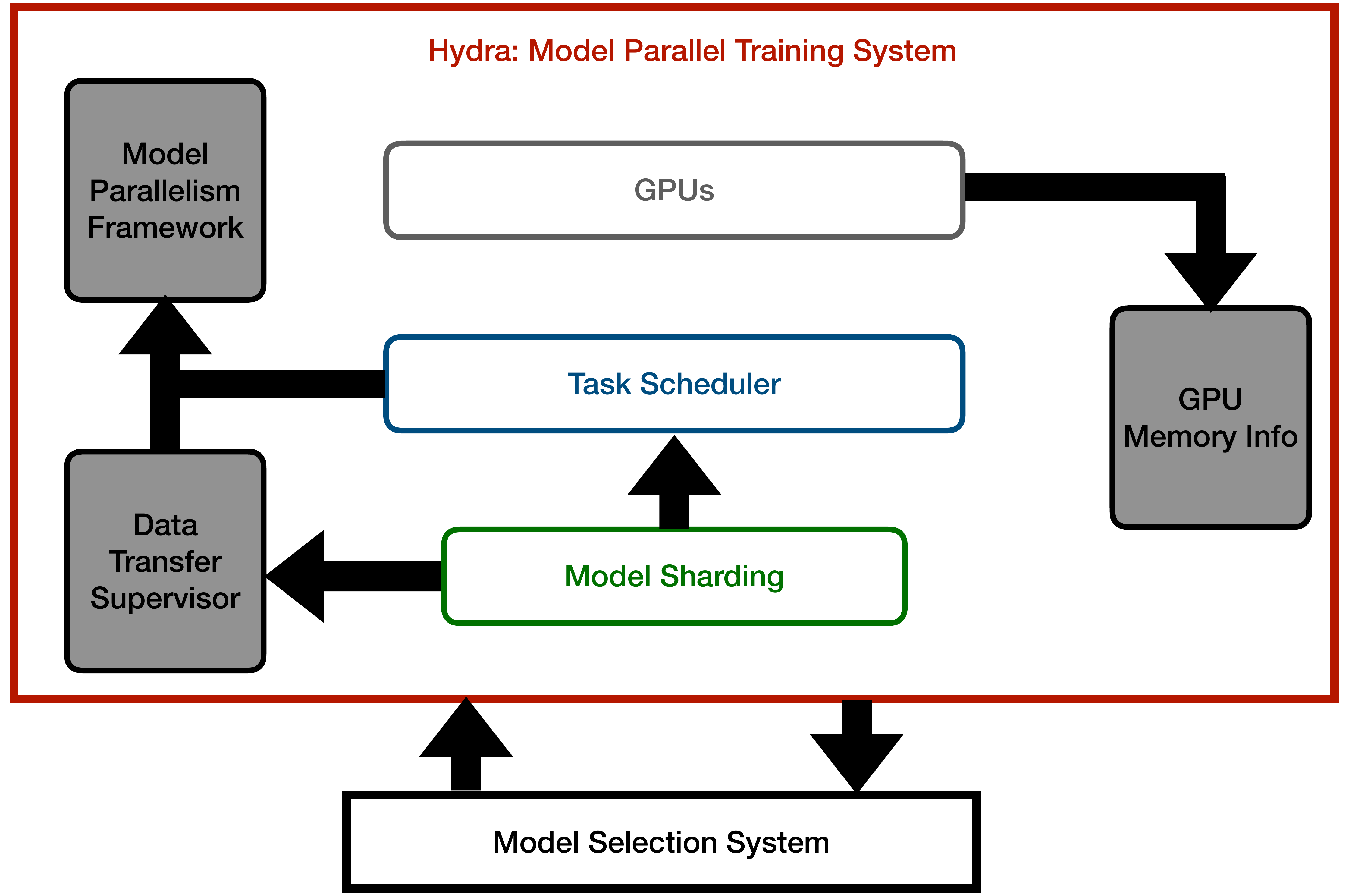}
  \caption{System Architecture of \textsc{Hydra}. For our purposes, we will use Cerebro as the paired model selection system.}
\end{figure}

\subsection{RESULTS}
We have been able to produce baselines using traditional model parallelism on our test workloads. On our heaviest test, BERT-Large, the use of traditional model parallelism provided a 3X reduction in per-device memory usage. These initial results provide a sample of the existing model-parallelism landscape against which we can compare \textsc{Hydra}'s performance. 
\section{RELATED WORK}
While model-parallel model selection is a heretofore unexplored field, there has been a great deal of work on model parallelism and model selection separately. \par

\textbf{Model Parallelism Frameworks.} DistBelief \cite{DistBelief} and FlexFlow \cite{jia2018data} aim to provide an efficient framework for model parallelization. We leverage this work in our own task scheduler and system design.

\textbf{Neural Graph Optimization.} TASO \cite{TASO} optimizes neural computational graphs via operator substitution, thus improving performance and reducing costs. TASO's basic concept of graph reduction meshes well with model parallel's aim of distributing costs. \par

\textbf{Integrated Parallelism.} Gholami et al. \cite{gholami2018integrated} investigated the benefits of integrating model parallelism with data parallelism. Similarly, we aim to integrate model parallelism into the task-parallel world of model selection.

\textbf{Optimized Model Selection.} Most existing model selection systems utilize some form of data or task parallelism. Google Vizier \cite{GoogleVizier} and Ray Tune \cite{liaw2018tune} use task parallelism for simultaneous model training. Cerebro \cite{inproceedings} combines task parallelism with data parallelism in a ``model-hopper" approach that bears some similarities to the blended task-model parallel approach we take with \textsc{Hydra}.

\section{TAKEAWAYS}
Model parallelism is a necessary tool to handle larger-than-memory models. Alas, the sequential nature of DL training has rendered device underutilization a major problem for model-parallel designs. Model selection, and indeed, multi-model training in general, offers a unique opportunity to utilize a more efficient version of model parallelism that does not suffer from these underutilization woes. In this paper, we present our plans for \textsc{Hydra}: our model-parallel model selection system. \textsc{Hydra}'s introduction of ``shard parallelism'' promises not only to democratize massive models, but also to boost training efficiency in all multiple model training regimes.

\bibliographystyle{unsrt}
\bibliography{SIGMOD_SRC21_MPMS}

\begin{thebibliography}{10}

\bibitem{devlin2019bert}
Jacob Devlin, Ming-Wei Chang, Kenton Lee, and Kristina Toutanova.
\newblock Bert: Pre-training of deep bidirectional transformers for language
  understanding, 2019.

\bibitem{turc2019}
Iulia Turc, Ming-Wei Chang, Kenton Lee, and Kristina Toutanova.
\newblock Well-read students learn better: On the importance of pre-training
  compact models.
\newblock {\em arXiv preprint arXiv:1908.08962v2}, 2019.

\bibitem{rajpurkar2016squad}
Pranav Rajpurkar, Jian Zhang, Konstantin Lopyrev, and Percy Liang.
\newblock Squad: 100,000+ questions for machine comprehension of text, 2016.

\bibitem{inproceedings}
Supun Nakandala, Yuhao Zhang, and Arun Kumar.
\newblock Cerebro: Efficient and reproducible model selection on deep learning
  systems.
\newblock pages 1--4, 06 2019.

\bibitem{DistBelief}
Jeffrey Dean, Greg~S. Corrado, Rajat Monga, Kai Chen, Matthieu Devin, Quoc~V.
  Le, Mark~Z. Mao, Marc'Aurelio Ranzato, Andrew Senior, Paul Tucker, Ke~Yang,
  and Andrew~Y. Ng.
\newblock Large scale distributed deep networks.
\newblock In {\em Proceedings of the 25th International Conference on Neural
  Information Processing Systems - Volume 1}, NIPS'12, page 1223–1231, Red
  Hook, NY, USA, 2012. Curran Associates Inc.

\bibitem{jia2018data}
Zhihao Jia, Matei Zaharia, and Alex Aiken.
\newblock Beyond data and model parallelism for deep neural networks, 2018.

\bibitem{TASO}
Zhihao Jia, Oded Padon, James Thomas, Todd Warszawski, Matei Zaharia, and Alex
  Aiken.
\newblock Taso: Optimizing deep learning computation with automatic generation
  of graph substitutions.
\newblock In {\em Proceedings of the 27th ACM Symposium on Operating Systems
  Principles}, SOSP '19, page 47–62, New York, NY, USA, 2019. Association for
  Computing Machinery.

\bibitem{gholami2018integrated}
Amir Gholami, Ariful Azad, Peter Jin, Kurt Keutzer, and Aydin Buluc.
\newblock Integrated model, batch and domain parallelism in training neural
  networks, 2018.

\bibitem{GoogleVizier}
Daniel Golovin, Benjamin Solnik, Subhodeep Moitra, Greg Kochanski, John Karro,
  and D.~Sculley.
\newblock Google vizier: A service for black-box optimization.
\newblock In {\em Proceedings of the 23rd ACM SIGKDD International Conference
  on Knowledge Discovery and Data Mining}, KDD '17, page 1487–1495, New York,
  NY, USA, 2017. Association for Computing Machinery.

\bibitem{liaw2018tune}
Richard Liaw, Eric Liang, Robert Nishihara, Philipp Moritz, Joseph~E. Gonzalez,
  and Ion Stoica.
\newblock Tune: A research platform for distributed model selection and
  training, 2018.

\end{thebibliography}
\end{document}